\documentclass[12pt]{iopart}

\usepackage{graphicx}
\usepackage{bm}

\begin{document}

\NJP

\title[Scattering Dominated Spatial Coherence and Phase Correlation Properties...]{Scattering Dominated Spatial Coherence and Phase Correlation Properties in Plasmonic Lattice Lasers}

\author{Janne I. Heikkinen$^1$, Benjamin Asamoah$^1$, Roman Calpe$^1$, Marek Ne{\v{c}}ada$^{1,2}$, Matias Koivurova$^3$ and Tommi K. Hakala$^1$}
\address{$^1$ Institute of Photonics, University of Eastern Finland}
\address{$^2$ Department of Applied Physics, Aalto University School of Science}
\address{$^3$ Faculty of Engineering and Natural Sciences, Photonics, Tampere University}
\ead{tommi.hakala@uef.fi}
\vspace{10pt}
\begin{indented}
\item[]September 2022
\end{indented}

\begin{abstract}
We present a comprehensive study of the polarization and spatial coherence properties of the lasing modes supported by a 4-fold symmetric plasmonic lattice. By modifying only, the scattering properties of the individual particles while keeping the lattice geometry constant, we are able to distinguish the scattering induced effects from the lattice geometry induced effects.  Customized interferometric measurements reveal that the lasing emission undergoes a drastic change from 1D to 2D spatial coherence with increasing particle size, accompanied with dramatic changes in the far field polarization and beaming properties. By utilizing T-matrix scattering simulations, we reveal the physical mechanism governing this transition. In particular, we find that there exists increased radiative coupling in the diagonal directions at the plane of the lattice when the particle diameter is increased. Finally, we demonstrate that the x- and y-polarized (degenerate) lasing modes become phase locked with sufficiently large particles.
\end{abstract}

%
%
%
%
%

\section{Introduction}

In plasmonic nanoparticle lattices, the radiative coupling of single particle plasmon resonances with the diffracted orders of the lattice give rise to hybrid modes known as surface lattice resonances (SLRs), which can possess particularly narrow linewidths \cite{zou2004silver,kravets2008extremely,auguie2008collective}. 
SLRs combined with strong near fields of localized plasmon resonances and molecular emitters have enabled coherent emission phenomena such as lasing and Bose-Einstein condensation in both weak \cite{hakala2017lasing,hakala2018bose} and strong coupling regimes \cite{vakevainen20,ramezani17}.
Lasing characteristics of plasmonic lattice lasers (PLLs) depend on the dispersion of SLR modes, which can be conveniently tuned by varying parameters such as lattice geometry, interparticle distance, as well as material, size, and shape of the constituent nanoparticles. \cite{humphrey14,guo17, yang14, knudson19, zhou13, fernandez22, heilmann22}. 
These various degrees of freedom indicate that many areas of the parameter space may still be unexplored, even in some of the simplest cases.
For instance, in the case of square lattice symmetry with cylindrical particles, the polarization dependence of the radiative feedback and the consequent anisotropy of the spatial coherence of lasing has only recently been recognized \cite{Asamoah_2021}.
The results from Ref.\cite{Asamoah_2021} suggest that since the feedback mechanism is based on radiative coupling, the directionality and the polarization dependence of the nanoparticle scattering govern the spatial coherence properties of the laser. Thus, the control over the scattering properties of the particles (and radiative feedback) could play a crucial role also for the beaming properties of these coherent light sources.
Various polarization and beaming patterns of PLLs have been observed by modifying the lattice geometry from the most simple (square) case towards more complex ones, for instance rectangular, hexagonal and honeycomb lattices \cite{Zhou2013, Hoang2017, Pourjamal2019, Tenner2018, Guo2019}. Here, we take an alternative route and instead of modifying the geometry of the lattice, we modify the single particle scattering properties while keeping the lattice symmetry (and other parameters of the lattice) constant. This allows us to distinguish the single particle scattering induced effects from the effects induced by the lattice geometry. Surprisingly, we observe a transition of spatial coherence properties from one to two dimensions in response to increasing particle diameter. The transition is accompanied with a drastic change in the polarization and beaming properties of the sample for both x- and y-polarizations. Further, we demonstrate a phase locking between degenerate x- and y-polarized lasing for large enough diameters. The physical mechanism governing both the transition of spatial coherence from one to two dimensions, as well as the phase locking of the x- and y-polarizations is identified by means of customized T-matrix scattering simulations. Our results demonstrate the crucial role of the size dependent scattering properties of the nanoparticles to the lasing characteristics of the PLL sources.


\section{Results}

\textbf{Sample preparation and optical setup.} The lattice consists of cylidrical gold nanoparticles with a period of $580$~nm in both x- and y-directions. The particles reside on a borosilicate substrate overlaid with a gain medium containing fluorescent IR-792 molecules in BA:DMSO (2:1) solution, whose refractive index matches the substrate index. The gain medium was pumped with a pulsed femtosecond laser (792 nm, 1 kHz, 150 fs). All the measurement results are  averaged over 300 pulses. We focused on particle diameters of $80$, $100$ and $120$~nm. The mode frequencies for each sample were obtained by measuring angle and wavelength resolved transmittances with a white light source, see Supporting Information Figs.~1 and 2.

\begin{figure}
    \centering
    \includegraphics[width=1\columnwidth]{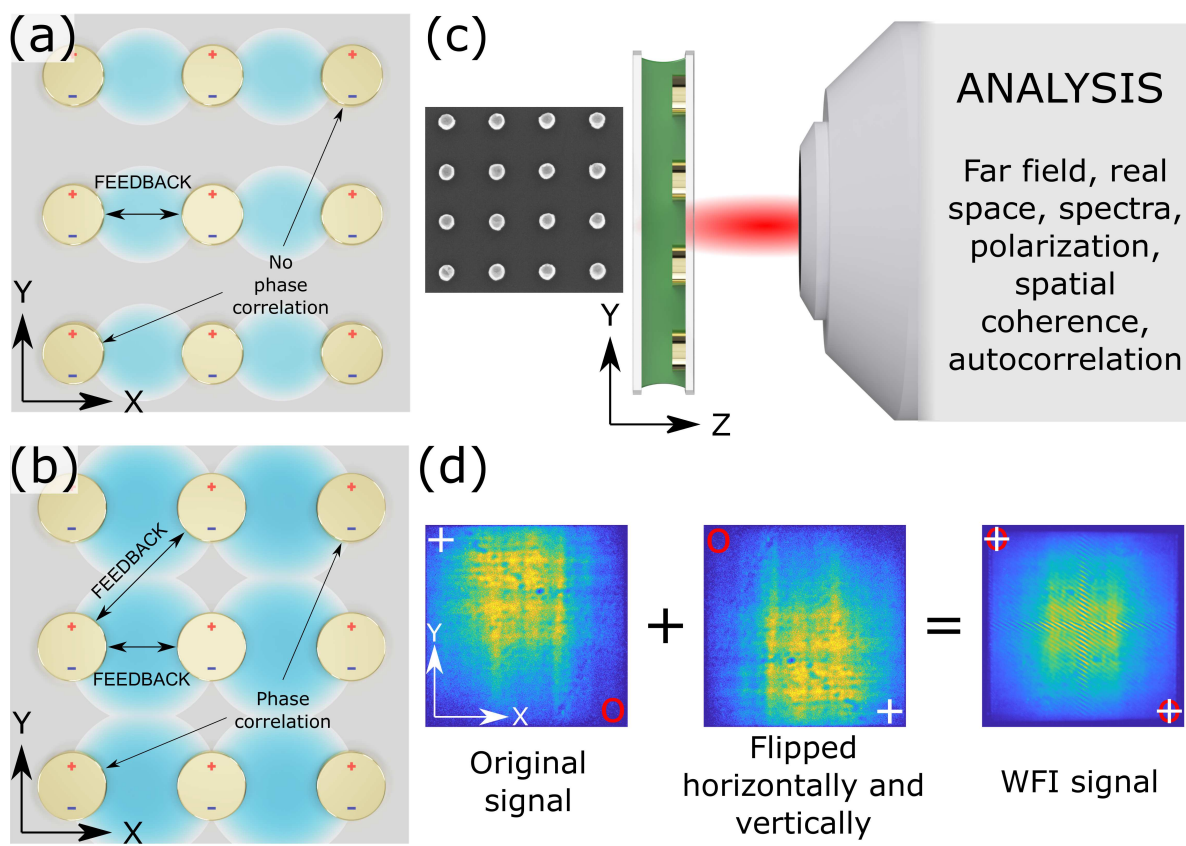}
    \caption{(a) Small particle diameters are expected to radiate similar to ideal dipoles such that y-polarized dipole moments radiate predominantly to x-direction. Thus, in the lasing regime, the radiative feedback is expected to take place in x-direction as well (as indicated by the black arrows). Consequently, the spatial coherence is expected to be high in x-direction. (b) Increasing the particle diameter modifies the scattering properties of the particle, potentially allowing for radiative feedback in both x- and y-directions. (c) The scheme for analysis and a SEM image of the particle lattice. (d) The WFI flips the source plane images with respect to the center ($x=0$, $y=0$) of the lattice, allowing for interference and spatial coherence measurements between any point pairs ($x$, $y$) and ($-x$, $-y$).}
    \label{fig:Fig1}
\end{figure}

Figs.~1 (a, b) summarize the main hypothesis of our manuscript. For conciseness, we consider only y-polarized dipoles in the figures. For small particles, the radiative feedback takes place only in x-direction, see Fig.~\ref{fig:Fig1} (a). Thus, spatial coherence is expected to be predominantly extended in x-direction as well.  Due to modified scattering properties, a larger particle diameter induces 1) a modified scattering pattern, allowing the feedback to occur also in diagonal directions of the lattice and 2) significantly higher overall scattering intensity, see Fig.~\ref{fig:Fig1} (b). Thus, a transition from one to two-dimensional spatial coherence may be feasible. Further, if the increased diameter increases the cross-coupling between x- and y-polarizations, the phase locking between otherwise independent x- and y-polarized degenerate lasing modes modes can take place.  
The measurement scheme and a SEM picture of the particle lattice are presented in figure \ref{fig:Fig1} (c). More detailed description of the setup is presented in the Supporting Information Fig.~1. 

We employ a custom built wavefront folding interferometer (WFI) \cite{koivurova2019scanning,halder2020mirror} to study the spatial coherence properties of the PLL emission. The two arms of the WFI flip the source plane image in horizontal (x) and vertical (y) directions, respectively, allowing us to map field correlations between spatial points $(x, y)$ and $(-x, -y)$ on the lattice. An example of this is shown in the figure \ref{fig:Fig1} (d).

%

\begin{figure}
    \centering
    \includegraphics[width=1\columnwidth]{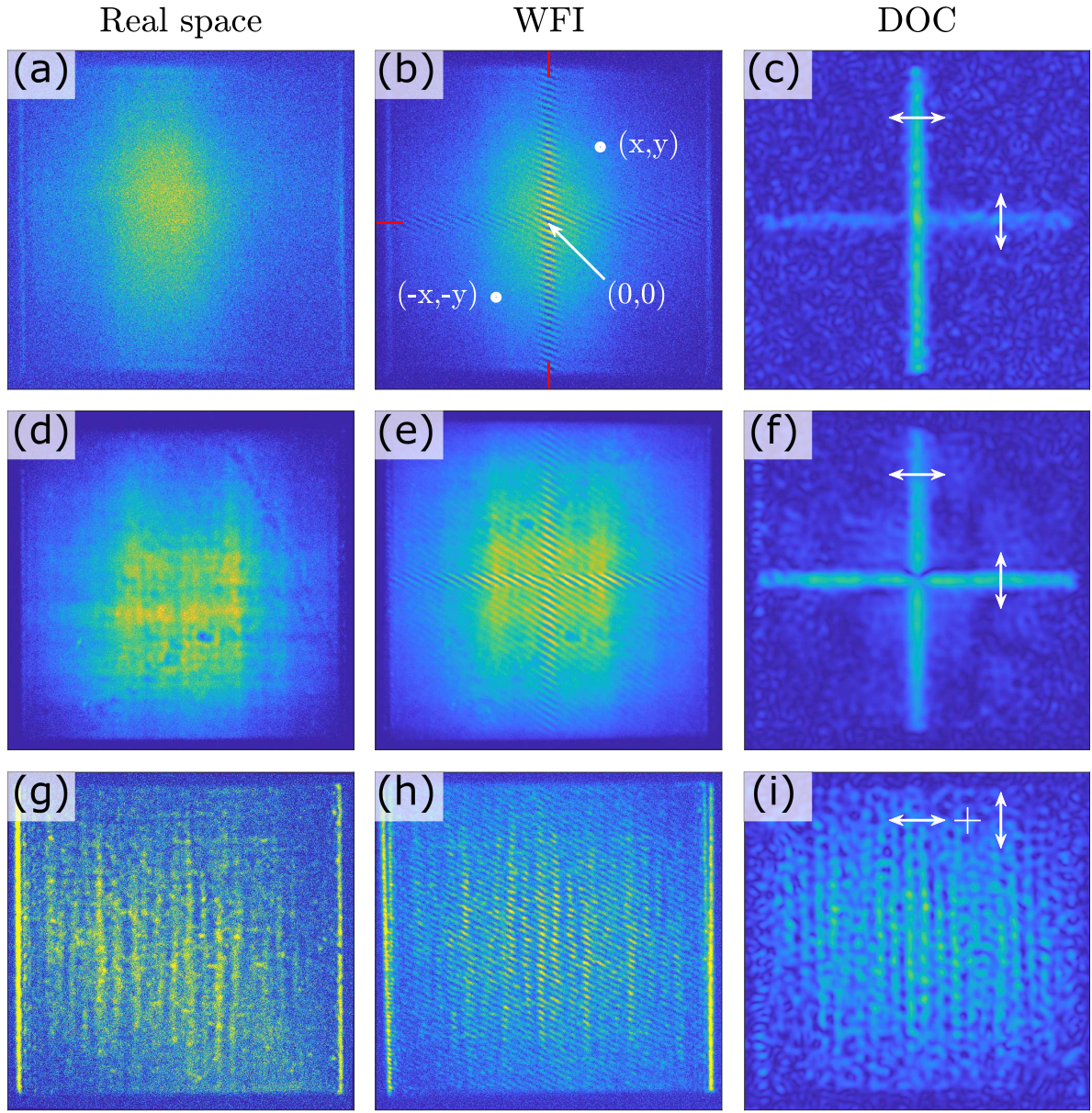}
    \caption{The obtained source plane data for 80 nm (a-c), 100 nm (d-f), and 120 nm (g-i) particles. The first column shows the source plane intensity, while the second and third show the WFI data and the obtained spatial degree of coherence (DOC), respectively. The white arrows indicate the observed polarizations in (c,f,i).}
    \label{fig:Fig2}
\end{figure}

\textbf{Measurement results.} In Fig. \ref{fig:Fig2} (a) is shown the real space intensity distribution for the 80 nm diameter particle array with a pump fluency of $1.1P_\mathrm{th}$. The lasing emission spectra as well as the threshold curves are shown in Supporting Information Figs.~2-5. 
First, we note that the center of the array is bright, with gradually decreasing intensity towards the edges. Previously, such intensity patterns have been associated with the so-called bright mode of the plasmonic lattice, a hybrid composed of diffracted orders of the lattice and dipolar plasmonic excitations in each particle \cite{hakala2017lasing}. The dipolar excitations of these collective SLR modes are weaker at the edges of the lattice due to reduced incident radiation on each particle when moved away from the lattice center. Somewhat similar intensity patterns are observed for Figs.~\ref{fig:Fig2}~(d, g) for 100 and 120 nm diameter particles respectively. Intriguingly, the 120 nm case shows more pronounced, high spatial frequency intensity variation in the center of the lattice.     


Figures \ref{fig:Fig2} (b), (e) and (h) contain the correlation information measured with the WFI. In Fig. \ref{fig:Fig2} (b) one can readily identify pronounced interference fringes along the two axes where $x=0$ and $y=0$ (marked by the red lines). Furthermore, the result implies negligible phase correlations elsewhere, namely between points $(x,y)$ and $(-x,-y)$, when both $x$ and $y$ are non-zero. Such a behaviour is indicative of 1-dimensional lasing feedback. For conciseness, we choose the $x=0$ fringes for closer inspection. The fringes extend over the whole lattice in the y direction, implying that the spatial coherence is limited by the lattice size. 
Strikingly, the fringes extend over a very short, approximately $5.8~\mu$m, or approximately $8$ particles in x-direction, indicative of very small spatial coherence. The degree of spatial coherence obtained from the fringe visibility is shown in the Fig. \ref{fig:Fig2} (c), exhibiting a cross pattern.
Polarization resolved analysis reveals that the two arms of the cross have orthogonal linear polarizations, as indicated by the arrows. 

With 100 nm diameter particles we observe qualitatively similar behaviour (second row of Fig. \ref{fig:Fig2}) as in the 80 nm case. However, a slightly larger area of emission is observed (Fig.~\ref{fig:Fig2} (d)). Further, the WFI signal reveals a somewhat more pronounced interference fringes in particular along the x-axis, see Fig.~\ref{fig:Fig2} (e). Notably, in this case the degree of coherence for both x- and y-polarizations are approximately equal (Fig.~\ref{fig:Fig2} (f)). 



As the particle diameter is increased to 120 nm, a drastic transition in the overall behaviour is observed, see Figs. \ref{fig:Fig2}  (g-i). In particular, the interference fringes are visible over the entire array, see Fig. 2 (h). This suggests well defined phase correlations and spatial coherence in 2 dimensions, such that even the points away from the axes $x = 0$ and $y = 0$ 
are phase correlated. The obtained degree of coherence in the Fig. \ref{fig:Fig2} (i) further confirms this conclusion: The degree of coherence is significant over the entire center part of the lattice with equal contributions from both x- and y-polarizations. 

\begin{figure}
    \centering
    \includegraphics[width=1\columnwidth]{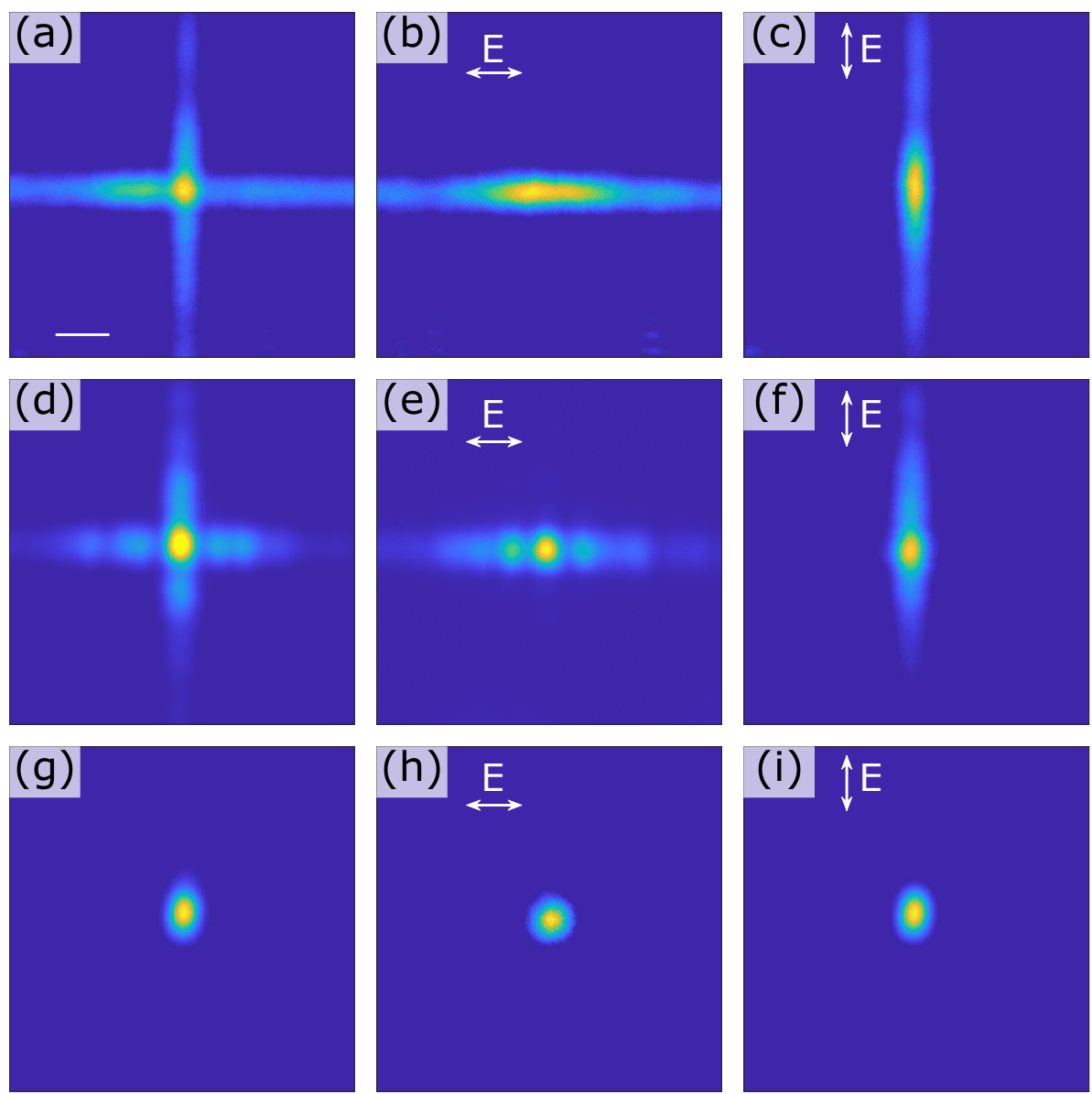}
    \caption{The obtained far field data for 80 nm (a-c), 100 nm (d-f), and 120 nm (g-i) particles. The first, second and third columns show the unpolarized, x- and y-polarized far field intensities, respectively. The scale bar in figure (a) is 1 degree. The scale is the same for all figures.}
    \label{fig:Fig3}
\end{figure}

Figure \ref{fig:Fig3} (a-i) shows the far field emission of the PLL, which was characterized by polarization resolved Fourier imaging.  For 80 nm diameter particles, a cross shaped intensity pattern is observed, see Fig. \ref{fig:Fig3} (a). The horizontal line of the cross consists of x-polarized and the vertical line of y-polarized light, see Figs. \ref{fig:Fig3} (b) and (c). Notably, the horizontal (vertical) line of the cross appears in the far field when the interference fringes in the source plane appear in the vertical (horizontal) direction. Thus, the features in the far field are localized in the direction in which the source plane spatial coherence extends over the entire lattice. Similar phase dependent behaviour was recently demonstrated in 2D plasmonic condensates. \cite{taskinen21} 
The angular divergences obtained from these images are $\delta \theta_x = 0.48^{\circ}, \delta \theta_y = 0.50^{\circ}$ for Fig.~3 (a), which, assuming a fully coherent beam gives a spatial coherence width of $\delta x = 2\times\pi/\delta k_x = 104~\mu\mathrm{m}$, and $\delta y = 2\times\pi/\delta k_y = 101~\mu\mathrm{m}$. The obtained numbers are in excellent accordance with the lattice dimensions ($100~\mu\mathrm{m} \times 100~\mu\mathrm{m}$).


For 100 nm diameter particles (middle row of Fig. \ref{fig:Fig3}) a somewhat similar far field pattern is observed, with the distinction that the cross shaped pattern is slightly more localized at $\theta_{x,y}=0$. This is expected due to the slightly larger area of spatial coherence compared to the 80 nm particle case, as seen when from the Figs. \ref{fig:Fig2} (c) and (f).

For 120 nm diameter, an entirely different far field pattern is observed, as seen in Figs. 3 (g-i). Importantly, the increased spatial coherence as seen in Fig. 2 dramatically reduces the asymmetry of the far-field distribution. Remarkably, only 20 nm increment in the particle diameter can induce this effect. 

\begin{figure}
    \centering
    \includegraphics[width=1\columnwidth]{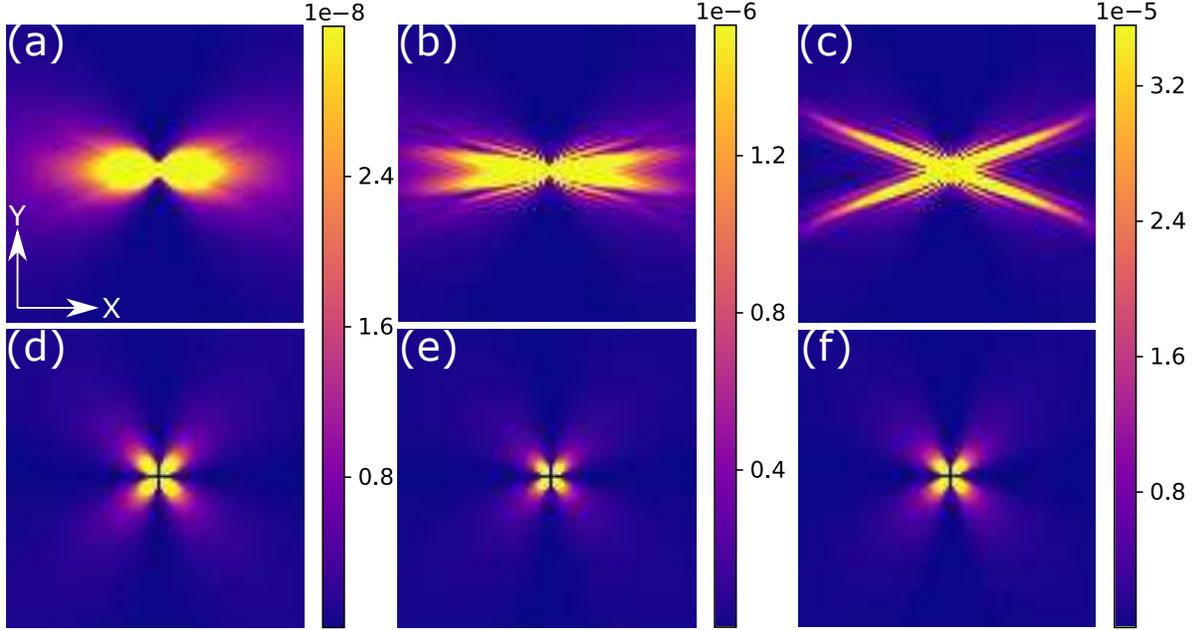}
    \caption{T-matrix scattering simulations for 80nm (a,d), 100 nm (b,e) and 120 nm (c,f) diameters. Top row shows y-polarized scattering intensities at the lattice plane, while the bottom row shows the x-polarized fields. In all cases, a single dipole in the center of the lattice is driven by a y-polarized dipole.}
    \label{fig:Fig4}
\end{figure}

\textbf{T-matrix simulations.} 
To rationalize these results, we carried out multiple scattering
T-matrix simulations including contributions from electric and magnetic dipoles, quadrupoles and hexapoles. The technical details of the method
can be found from our previous work \cite{Marek2021}.  For conciseness,
we present the results for the case where a single
nanoparticle in the center of the lattice is driven with a y-polarized
dipole spherical wave, see Fig.~\ref{fig:Fig4}. Note that due to symmetry, analogous results would be
obtained for the x-polarized case. Nominal lattice periodicity, size
and shape of the particles, as well as dielectric functions of the
materials are the same as in the experiments. The first observation from
the simulations are that the scattering pattern of y-polarized fields
changes drastically with increasing particle diameter, see Figs.~4
(a-c). For 80 nm particles the pattern appears almost like a dipolar
radiation pattern: the maximum intensity is observed in the x-direction
of the driven particle. However, for the 100 nm case, the maxima reside
slightly away from the x-direction. For 120 nm, the maxima form a
cross-shaped pattern, with almost negligible intensity in the
x-direction. Another noteworthy observation is that the overall
scattering intensity increases by over 3 orders of magnitude when
increasing the diameter from 80 to 120 nm. 
We believe that the drastic
changes observed in the spatial coherence properties in Fig.~2 and the
corresponding changes in the beaming properties in Fig.~3 are due to
both modified scattering pattern as well as increased scattering
intensity for large enough particles. Due to highly nonlinear character
of stimulated emission processes in plasmonic nanolasers, even a minor
variation in scattering efficiency can induce sufficient radiative
feedback and result in macroscopic phase coherence over the entire
lattice.

To summarize, our results in Figs. 2 and 3 established an intimate connection between the particle diameter and the spatial coherence and far field beaming properties of plasmonic lasers. The most notable changes occurred between 100 nm and 120 nm particles, with the phase correlations extending throughout the entire center part of the lattice for the 120 nm case. This suggests that the degree of coherence underwent a transition from one to two dimensions for both x- and y-polarizations. Fig.~\ref{fig:Fig4} revealed the underlying physical mechanism for the drastic changes observed in the experiments, namely the modified radiation pattern as well as increased radiation intensity with large diameter particles. An interesting question then arises, whether there also exist well defined phase correlations between x- and y-polarizations?

In Figs.~4 (d-f), we present the scattering intensities for x-polarized fields, while still driving the central particle with y-polarization. Notably, while the pattern stays qualitatively the same for all particle sizes, the intensity exhibits similar increment (3 orders of magnitude) with increasing particle diameter. Thus, for large enough particles, y-polarized dipoles could indeed produce sufficient x-polarized scattered fields to establish phase correlations between x- and y-polarizations of the lasing signal. Due to the symmetry of the lattice, the x- and y-polarized modes are degenerate, and thus the phase correlations can be conveniently studied by measuring the Stokes parameters.

\textbf{Stokes parameters.}
To study the potential correlations between x- and y-polarizations, we carried out an experiment to recover the Stokes parameters of the far field radiation for 100 nm and 120 nm particle sizes, see Fig. 5. For the details of the experiment, see Supporting Information Fig. 1.
Fig. 5 (a) presents the S1 parameter for d = 100 nm sample. Note, that S1 describes the  relative fraction of x- and y-polarized light. Value S1 = 1 implies x-polarized light, S1 = -1 implies y polarized light and S1 = 0 indicates equal contributions from both x- and y-polarized light. Notably, the Stokes parameter measurements are in full agreement with the polarization resolved far field analysis in Fig. 3: The vertical arm of the emission is y-polarized, while horizontal arm is x-polarized. The overlap region of both arms results in S1 = 0, indicating equal contributions from x- and y-polarizations. The Stokes parameter S2 describes the degree of diagonal polarization, and the value S2 = 0 in Fig. 5 (b) indicates no diagonal polarization. Further, S3 = 0 in Fig. 5 (c) indicates no circular polarization is present. Altogether, the results for 100 nm diameter suggest that each arm of the cross has a linear polarization and in the overlap region, the x- and y-polarizations have no well-defined phase difference (i.e., they are not phase locked). For 120 nm sample the S1 parameter exhibits a different distribution, see Fig. 5 (d). In particular, within the area of the main beam (indicated by the circle), S1 is approximately zero. This suggests equal contributions from x- and y-polarizations throughout the beam area. More importantly, the S2 = -1, suggesting a -45 degree polarization and therefore a well-defined phase correlation between x- and y-polarizations. This is in stark contrast to 100 nm case, where S2 = 0  and no phase correlations are present.  Our observations can be explained by the comparison between Figs.~4 (e, f). While in both figures the x-induced dipoles have a similar pattern, for 120 nm particles the intensity of the x-induced dipole is over 20 times higher than for 100 nm particle. Thus, a large particle diameter enables not only 2-dimensional spatial coherence for both polarizations (as observed in Figs. 2 and 3), but it also enables phase correlations between two orthogonal polarizations. While the detailed study of the required conditions for such phase locking remains a scope for future work, some considerations can nevertheless be put forward. In our previous work \cite{Asamoah_2022}, we have shown that under the same exact experimental conditions (including the pump, particle shape, material, lattice periodicity), a further increase of the diameter to 140 nm produces lasing from so-called bound state in continuum (BIC) modes, whose origin lies in the quadrupolar resonances of large diameter particles. It is feasible that  the 120 nm is a border line case, where the multipolar contribution to the resonance is sufficient to phase lock the x- and y-polarizations. 

\begin{figure}
    \centering
    \includegraphics[width=1\columnwidth]{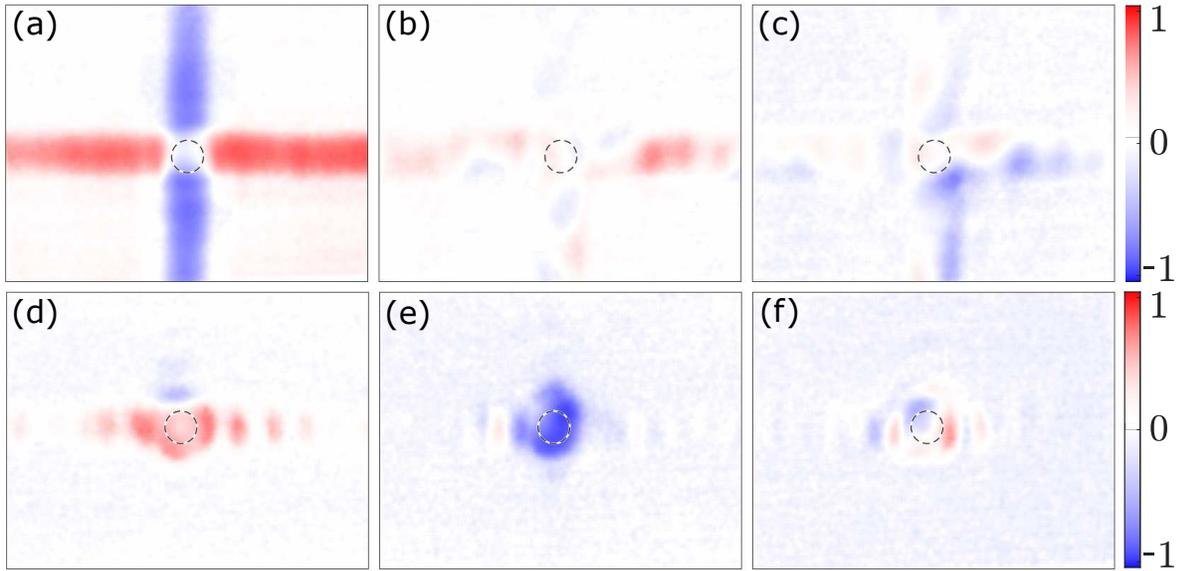}
    \caption{The Stokes parameters determined from far field emission for 100 nm (a, b and c) and 120 nm particles (d, e and f). First, second and third column show the S1, S2 and S3 parameters, respectively. The dashed lines indicate the FWHM intensities of the unpolarized beams.}
    \label{fig:Fig5}
\end{figure}
\section{Conclusions}

To conclude, we have studied the single particle scattering induced effects in lasing plasmonic lattices. By keeping the lattice geometry constant and varying only the particle diameter we were able to distinguish lattice geometry induced effects from single particle scattering induced effects. For the first time, we demonstrate a transition of the spatial coherence from one to two dimensions with increasing particle diameter. The far field emission undergoes a transition from a cross-shaped pattern to a symmetric, approximately circular beam in response to changes in the spatial coherence. T-matrix simulations indicate that the physical mechanism governing this transition is associated with the increased scattering to the diagonal directions in the lattice. The symmetry of the lattice allows the coexistence of two independent, orthogonally polarized modes. Strikingly, the cross-coupling of y-polarized dipole radiation to x-polarization is strongly increasing with particle size. With large enough diameters, this induces a well-defined phase correlation in otherwise independent (but degenerate) x- and y-polarized modes. 





\section{Methods}

\textbf{Correlation functions.} Here we present the main equations necessary for the analysis of our experimental results. The emission considered in this work is pulsed with a picosecond-scale pulse length, with spectral linewidths on the nanometer scale. The emission may be anisotropic in the sense that the divergence properties in the x- and y-directions can be different with varying particle diameter.

The electric field generated by the source is denoted by $E(\boldsymbol{\rho};t)$, where $\boldsymbol{\rho} = (x,y)$ contains the transverse coordinates, and $t$ is the time in the moving reference frame of the pulse. To quantify the spatial correlation properties of the field, we employ the time integrated mutual coherence function (MCF), defined as  
\begin{equation*}
    \Gamma\left(\boldsymbol{\rho}_1,\boldsymbol{\rho}_2\right) = 
    \int_{-\infty}^{\infty} \langle
    E^*(\boldsymbol{\rho}_1;t)E(\boldsymbol{\rho}_2;t) \rangle
    \mathrm{d}t,
\end{equation*}
where the angle brackets denote ensemble averaging. This form of the MCF is the relevant one when performing measurements with a slow detector. Moreover, we have set the time delay between the two copies of the field to zero, since we are interested only on the spatial coherence properties.

If we set $\boldsymbol{\rho}_1 = \boldsymbol{\rho}_2 = \boldsymbol{\rho}$, the time integrated MCF yields the spatial intensity distribution, i.e. $\Gamma(\boldsymbol{\rho},\boldsymbol{\rho}) = I(\boldsymbol{\rho})$. The intensity can be used to define a normalized quantity
\begin{equation*}
    \gamma(\boldsymbol{\rho}_1,\boldsymbol{\rho}_2) = \frac{\Gamma(\boldsymbol{\rho}_1,\boldsymbol{\rho}_2)}{\sqrt{I(\boldsymbol{\rho}_1)I(\boldsymbol{\rho}_2)}},
\end{equation*}
which is a measure of the complex degree of spatial coherence of the time-averaged pulsed field. Then, the magnitude of the degree of coherence is obtained as DOC = $|\gamma|$.

\textbf{T-matrix simulations.} 
The propagation of electromagnetic fields inside the array were
simulated using the multiple-scattering T-matrix method implemented
by the open-source QPMS suite\cite{QPMS,Marek2021}. The particle in the
center of the array was excited with a y-polarized regular
electric dipole spherical wave (artificially local, i.e. its
direct effects are limited to that single particle) of unit intensity;
from that particle, the field was then allowed to scatter throughout
the whole array, yielding the patterns in Fig.~\ref{fig:Fig4}.

The simulation frequencies for each particle size were pre-determined
using mode calculations with corresponding \emph{infinite} periodic arrays\cite{Marek2021},
with the imaginary part of the mode frequency being discarded.
 
The electric permittivity of the metal was modelled using the Lorentz-Drude
formula with parameters taken from \cite{rakic_optical_1998}, the background
medium was set to have constant relative permittivity of 1.52.



\section{Acknowledgements}
We acknowledge Academy of Finland Flagship Programme, Photonics Research and Innovation \textrm{PREIN} 320165, 320166 and Academy of Finland project number 322002. We acknowledge the computational resources provided by the Aalto Science-IT project.

\section{References}
\bibliographystyle{unsrt}
\bibliography{main}

\end{document}